\documentclass[twocolumn,english,aps,prl,showpacs]{revtex4}
\usepackage{amsmath,graphicx,amssymb,epsfig,babel,dsfont,color}

\renewcommand{\Re}{{\rm Re}}
\renewcommand{\Im}{{\rm Im}}
\newcommand{\Tr}{{\rm Tr}}
\newcommand{\rd}{{\rm d}}

\newcommand{\kb}{k_{\rm B}}

\newcommand{\ro}{{\rm o}}
\newcommand{\re}{{\rm e}}
\newcommand{\ri}{{\rm i}}

\begin{document}

\title{Hyperbolic blackbody}

\author{Svend-Age Biehs$^{1}$}
\email{s.age.biehs@uni-oldenburg.de}
\author{Slawa Lang$^{2}$, Alexander Yu. Petrov$^{2,3}$, Manfred Eich$^{2}$}
\author{Philippe Ben-Abdallah$^{4}$}
\email{pba@institutoptique.fr}
\affiliation{$^1$Institut f\"{u}r Physik, Carl von Ossietzky Universit\"{a}t, D-26111 Oldenburg, Germany.}
\affiliation{$^2$ Institute of Optical and Electronic Materials, Hamburg University of Technology, 21073 Hamburg, Germany}
\affiliation{$^3$ ITMO University, 49 Kronverskii Ave., 197101, St. Petersburg, Russia.}
\affiliation{$^4$ Laboratoire Charles Fabry,UMR 8501, Institut d'Optique, CNRS, Universit\'{e} Paris-Sud 11,
2, Avenue Augustin Fresnel, 91127 Palaiseau Cedex, France.}

\date{\today}


\begin{abstract}
The blackbody theory is revisited in the case of thermal electromagnetic fields inside uniaxial anisotropic media 
in thermal equilibrium with a heat bath. When these media are hyperbolic, we show that the spectral energy density 
of these fields radically differs from that predicted by Planck's blackbody theory. We demonstrate that 
the maximum of their spectral energy density is shifted towards frequencies smaller than Wien's frequency 
making these media apparently colder. Finally, we derive Stefan-Boltzmann's law for hyperbolic media 
which becomes a quadratic function of the heat bath temperature.

\end{abstract}

\maketitle

In $1901$ Planck~\cite{Planck} derived the famous law describing the spectral distribution of energy 
of a blackbody (BB) by introducing the concept of quantum of light laying so the foundation of quantum physics. 
In his description of the problem, the electromagnetic field inside a cavity made with opaque 
walls which is set at a constant temperature is studied. In this formulation~\cite{Kirchoff}, the cavity is at 
thermal equilibrium and acts as a heat bath. The walls of the cavity emit and absorb electromagnetic waves so 
that the field itself becomes equilibrated. The internal energy density of the electromagnetic field in the cavity with volume $V$ 
for both principal polarization states (abbreviated by $\rm s$ and $\rm p$) is then given by
\begin{equation}
  U_{\rm BB}^{\rm s/p} = \frac{1}{2}\int_0^\infty\!\!\!\rd \omega\, \frac{\omega^2}{\pi^2 c^3}\frac{\hbar \omega}{e^{\frac{\hbar \omega}{\kb T}}-1} 
                       = \frac{\Gamma(4) \zeta(4)}{2\pi^2 }\frac{(\kb T)^4}{(\hbar c)^3},
  \label{Planck_law}
\end{equation}
where $\hbar$, $\kb$ and $c$ denote Planck's constant, Boltzmann's constant and the velocity of light 
in vacuum, while $\Gamma$ and $\zeta$ are Riemann's gamma and zeta functions. Here and in 
the following we neglect vacuum fluctuations.

%
%
%
%
\begin{figure}[Hhb]
  \epsfig{file =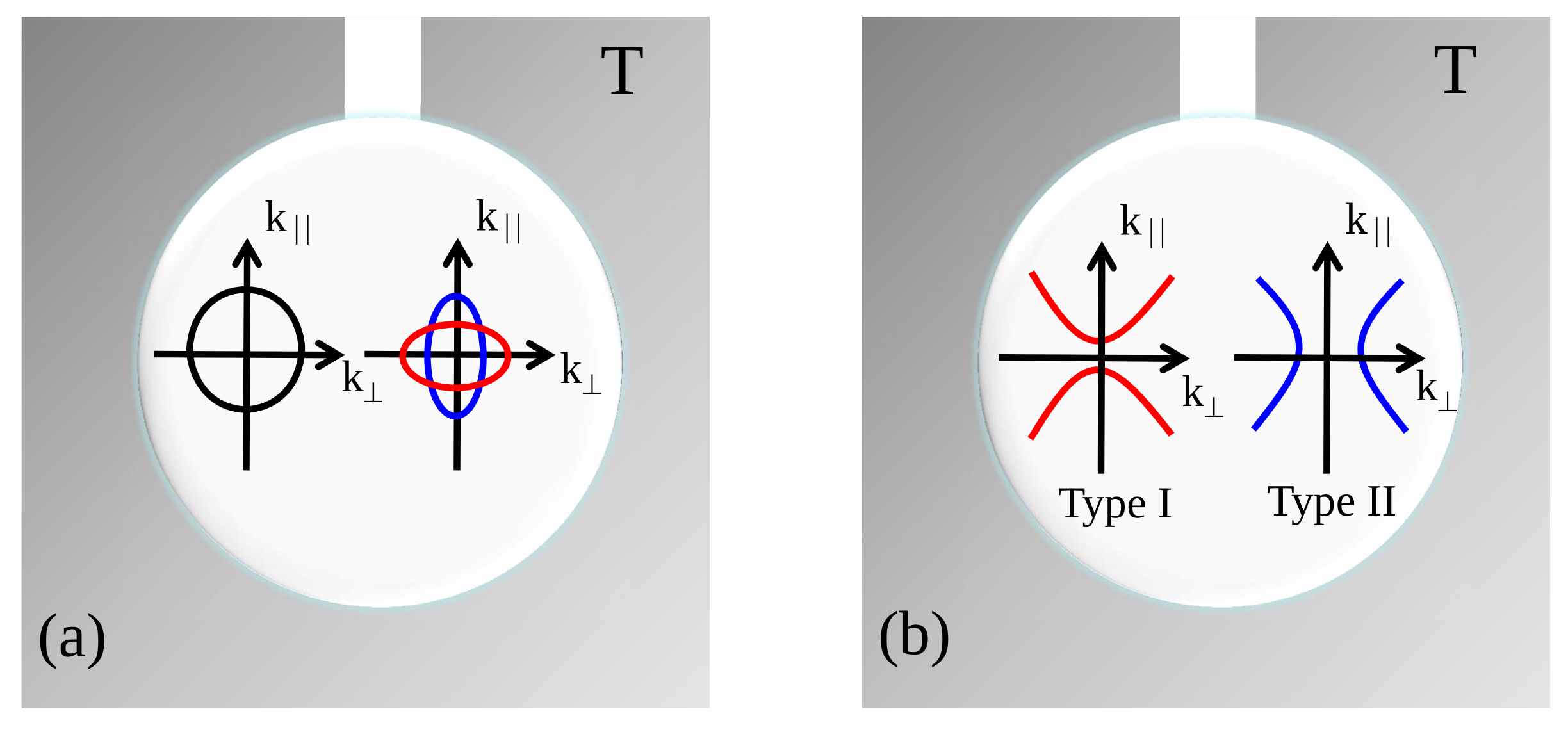, width = 0.5\textwidth}
  \caption{(a) Cavity at temperature $T$ containing an isotropic medium of permittivity $\epsilon > 0$ or an anisotropic (uniaxial) medium with 
               $\epsilon_\perp > 0$  and $\epsilon_\parallel > 0$. The particular case  $\epsilon_\perp =\epsilon_\parallel =1$ corresponds to the 
               {\itshape classical} BB. 
           (b) Cavity at temperature $T$ containing a hyperbolic medium of type  I ($\epsilon_\perp > 0$ and $\epsilon_\parallel < 0$) and of 
               type II ($\epsilon_\perp < 0$ and $\epsilon_\parallel > 0$). The isofrequency curves are plotted inside the cavities in 
               the plane ($k_\perp,k_\parallel$).}
  \label{Fig1}
\end{figure}

In this Letter, we revisit this old problem when the cavity is filled with a uniaxial medium ~\cite{YehBook}
with a relative permittivity tensor of the form
\begin{equation}
  \boldsymbol{\epsilon} = \begin{pmatrix} \epsilon_\perp & 0 & 0\\ 0 & \epsilon_\perp & 0\\ 0 & 0 & \epsilon_\parallel\end{pmatrix}.
\end{equation}
Here without loss of generality we assume that the optical axis points into the z-direction; 
$\epsilon_\parallel$ is the permittivity along the optical axis and $\epsilon_\perp$
is the permittivity perpendicular to the optical axis. For convenience, if not specified differently we neglect
dispersion, dissipation and nonlocal effects in the following. Within such materials so-called ordinary modes (OMs) and 
extraordinary modes (EMs) exist and satisfy the dispersion relations~\cite{YehBook}
\begin{align}
  \frac{k_\perp^2}{\epsilon_\perp} + \frac{k_\parallel^2}{\epsilon_\perp} &=  \frac{\omega^2}{c^2}, \qquad \text{(OM)} \label{Eq:OM} \\
  \frac{k_\perp^2}{\epsilon_\parallel} + \frac{k_\parallel^2}{\epsilon_\perp} &=  \frac{\omega^2}{c^2}, \qquad \text{(EM)} \label{Eq:EOM}
\end{align}
where $k_\perp$ ($k_\parallel$) is the wave number component perpendicular (parallel) to the optical axis. 
In usual dielectric uniaxial media the principal constants $\epsilon_\perp$ and $\epsilon_\parallel$ are both positive and 
the iso-frequency surfaces defined by relations (\ref{Eq:OM}) and (\ref{Eq:EOM}) are spheres or ellipsoids, resp., 
as illustrated in Fig.~\ref{Fig1}(a). On the other hand, when $\epsilon_\parallel < 0$ and $\epsilon_\perp > 0$ or 
$\epsilon_\parallel > 0$ and $\epsilon_\perp < 0$ the iso-frequency surfaces of the EM are two- or one-sheeted hyperboloids 
[see Fig.~\ref{Fig1}(b)]. The first class of such uniaxial medium is called hyperbolic of type I while the second one hyperbolic 
of type II~\cite{Smith2000,SmithSchurig2003}. Of course both $\epsilon_\perp$ and $\epsilon_\parallel$ can also be negative.
In such uniaxial metallic-like materials no propagating modes exist and the upcoming quantities are all 0.

To start, we focus our attention on the electromagnetic field inside a cavity filled with an arbitrary uniaxial 
medium. The spectral density of states (DOS) defined as the energy density $U$ normalized to the mean energy of a harmonic oscillator 
and associated to the thermal field can be calculated either by counting the modes in the wavevector space using 
expressions (3) and (4) or by means of the generalized trace formula ~\cite{MandelWolfBook,Novotny2012,SupplMat}
\begin{equation}
  D(\omega) = \frac{\omega}{c^2 \pi} \Im \Tr\bigl[ \boldsymbol{\epsilon} \mathds{G}^{\rm EE}(\mathbf{r,r},\omega) 
              + \boldsymbol{\mu} \mathds{G}^{\rm HH}(\mathbf{r,r},\omega) \bigr],
	\label{Eq:TraceDOS}
\end{equation}
where $\mathds{G}^{\rm EE}$ and $ \mathds{G}^{\rm HH}$ are the electric and magnetic Green's dyadics for the bulk material and
$\boldsymbol{\mu}$ is the relative permeability tensor. The result for a classical BB can be retrieved by using
$\boldsymbol{\epsilon} = \boldsymbol{\mu} = \mathds{1}$ with the unit dyad $\mathds{1}$. Then 
the above expression reduces to the well-known expression $D_{\rm BB}^{\rm s+p}(\omega)=\frac{\omega^2}{\pi^2 c^3}$.
In the following, for the sake of clarity, we consider non-magnetic materials (i.e. $\boldsymbol{\mu} = \mathds{1}$) only. 
By inserting the general expression of dyadic Green's tensors of uniaxial materials~\cite{Weiglhofer1990} in the trace 
formula (\ref{Eq:TraceDOS}) it is straight forward to derive the DOS for the three different classes of 
uniaxial media. Assuming that those media are lossless then in dielectric anisotropic media the DOS $D^\ro_{\rm D}$ for the OMs 
and $ D^\re_{\rm D}$ for the EMs are given by the following expressions~\cite{Eckhardt1978,SupplMat} 
\begin{equation}
  D^\ro_{\rm D}(\omega) = \frac{\omega^2}{\pi^2 c^3} \frac{\epsilon_\perp \sqrt{\epsilon_\perp}}{2}  \quad\text{and} \quad 
  D^\re_{\rm D}(\omega) = \frac{\omega^2}{\pi^2 c^3} \frac{\epsilon_\parallel \sqrt{\epsilon_\perp}}{2}. 
\end{equation}
On the other hand, in the hyperbolic case we obtain  
\begin{align}
  D^\ro_{\rm I} &= \frac{\omega^2}{\pi^2 c^3} \frac{\epsilon_\perp \sqrt{\epsilon_\perp}}{2}, \\
  D^\re_{\rm I} &= \frac{\omega}{\pi^2 c^2} \frac{|\epsilon_\parallel|}{2} \biggl(\sqrt{k_{\perp, {\rm max}}^2 \frac{\epsilon_\perp}{|\epsilon_\parallel|} + \frac{\omega^2}{c^2} \epsilon_\perp } - \frac{\omega}{c} \sqrt{\epsilon_\perp}\biggr)
\end{align}
and 
\begin{align}
  D^\ro_{\rm II}  &= 0, \\
  D^\re_{\rm II} &= \frac{\omega}{\pi^2 c^2} \frac{\epsilon_\parallel}{2} \sqrt{k_{\perp, {\rm max}}^2 \frac{|\epsilon_\perp|}{\epsilon_\parallel}- \frac{\omega^2}{c^2} |\epsilon_\perp| },
\end{align}
for the type I and type II media, respectively. Note that, we have introduced a cutoff wavenumber $k_{\perp,{\rm max}}$ which  
for dispersive media can be a function of the frequency and which is determined by the concrete (atomic or meta) structure 
of the medium. For an ideal hyperbolic 
material $k_{\perp,{\rm max}}$ is infinity so that the DOS diverges as was pointed out previously~\cite{SmolyaninovNarimanov2010}. However, 
for any real structure $k_{\perp,{\rm max}}$ is a finite quantity~\cite{DrachevEtAl2013}. For artificial hyperbolic
structures it is mainly determined by the unit-cell size of the meta structure. Note further that the DOS of the EMs of 
type I and type II hyperbolic media coincides for 
$k_{\perp,\rm max} \gg \frac{\omega}{c} \sqrt{|\epsilon_\parallel|}$ and is given by
\begin{equation}
  D^\re_{\rm I} \approx D^\re_{\rm II} \approx \frac{\omega}{\pi^2 c^2} \frac{\sqrt{|\epsilon_\perp \epsilon_\parallel|}}{2} k_{\perp,{\rm max}}.
\end{equation}

With the help of the DOS we can determine the thermodynamic potentials of the photon gas inside the uniaxial material. By definition,
the internal and the free energy per unit volume are given by~\cite{Landau}
\begin{align}
  U &= \int_0^\infty\!\!\rd \omega\, D(\omega) \mathcal{U}(\omega,T), \\ 
  F &= \int_0^\infty\!\!\rd \omega\, D(\omega) \mathcal{F}(\omega,T) 
\end{align}
where
\begin{align}
  \mathcal{U}(\omega,T) &= \frac{\hbar \omega}{\re^{\frac{\hbar \omega}{\kb T}} - 1}, \\
  \mathcal{F}(\omega,T) &= \kb T \ln\bigl(1 - \re^{-\frac{\hbar \omega}{\kb T}}\bigr).
\end{align}
Finally, from the internal and free energy we can also determine the entropy per unit volume by 
\begin{equation}
   S = \frac{U - F}{T}.
\end{equation}
Clearly, by means of these expressions we can derive any thermodynamic property of the photon
gas inside the cavity as the pressure $P=\frac{U}{3}$, the photonic heat capacity $C_{\rm V}=\frac{\partial U}{\partial T}$, etc. 

Let us first have a look at the expressions for the ordinary uniaxial material.
In this case we obtain 
\begin{equation}
  U^{\ro}_{\rm D}= U_{\rm BB}^{\rm s} \epsilon_\perp \sqrt{\epsilon_\perp} \quad \text{and} \quad U^{\re}_{\rm D} = U_{\rm BB}^{\rm p} \epsilon_\parallel \sqrt{\epsilon_\perp}.
\end{equation} 
Therefore, when $\epsilon_\perp = \epsilon_\parallel = 1$ we recover the classical blackbordy result. 
And the relation between the internal energy, the free energy and the entropy have the familiar forms
\begin{equation}
  F^{\ro/\re}_{\rm D} = - \frac{1}{3} U^{\ro/\re}_{\rm D} \quad \text{and} \quad  S^{\ro/\re}_{\rm D} = \frac{4}{3} \frac{U^{\ro/\re}_{\rm D}}{T}.
\end{equation}
Note that these relations are the same as for a usual BB because the DOS of the field inside a dielectric uniaxial medium 
is proportional to $\omega^2$.  

On the contrary, in type I and type II hyperbolic media we have seen  that the DOS of the EMs 
is linear in $\omega$ as in a 2D photon gas in vacuum. It follows that the relations between the thermodynamic properties of 
the photon gas are radically different in that case. Indeed, we 
obtain ($k_{\perp, \rm max} \gg \frac{\omega}{c} \sqrt{|\epsilon_\parallel|}$)
\begin{equation}
  U^\re_{\rm I/II} \approx \frac{\sqrt{|\epsilon_\parallel \epsilon_\perp|}}{2} k_{\perp,{\rm max}} \frac{1}{\pi^2 c^2} \frac{\Gamma(3) \zeta(3)}{\hbar^2 }(\kb T)^3
\label{Eq:EnergyDensityHyperbolic}
\end{equation}
and
\begin{equation}
  F^\re_{\rm I/II} = -\frac{1}{2} U^\re_{\rm I/II} \quad \text{und} \quad  S^{\re}_{\rm I/II} = \frac{3}{2} \frac{U^\re_{\rm I/II}}{T}
\end{equation}
Hence $U$, $F$ and $S$ are proportional to $T^3$ and not anymore to $T^4$. This result is a direct consequence of 
the linear dependence of the electromagnetic DOS inside hyperbolic media with respect to $\omega$. 
Naturally, for the OMs we find
\begin{equation}
   U^\ro_{\rm I} = U^\ro_{\rm D}\quad \text{and} \quad U^\ro_{\rm II} = 0.
\end{equation}
Note that for the type II hyperbolic material the internal energy of the OMs is zero, since
there are no OMs in such a material. The internal energy of the OMs in a type I
hyperbolic materials is just the same as in a dielectric uniaxial medium. Hence, the relations
between the thermodynamic potentials are the same as in a dielectric uniaxial medium. However, in typical hyperbolic (meta)materials the maximal wave vector is much larger than the vacuum wave vector $k_{\perp, \rm max} \gg \frac{\omega}{c}$ making the material properties dominated by the EMs.

\begin{figure}[Hhbt]
  \epsfig{file = 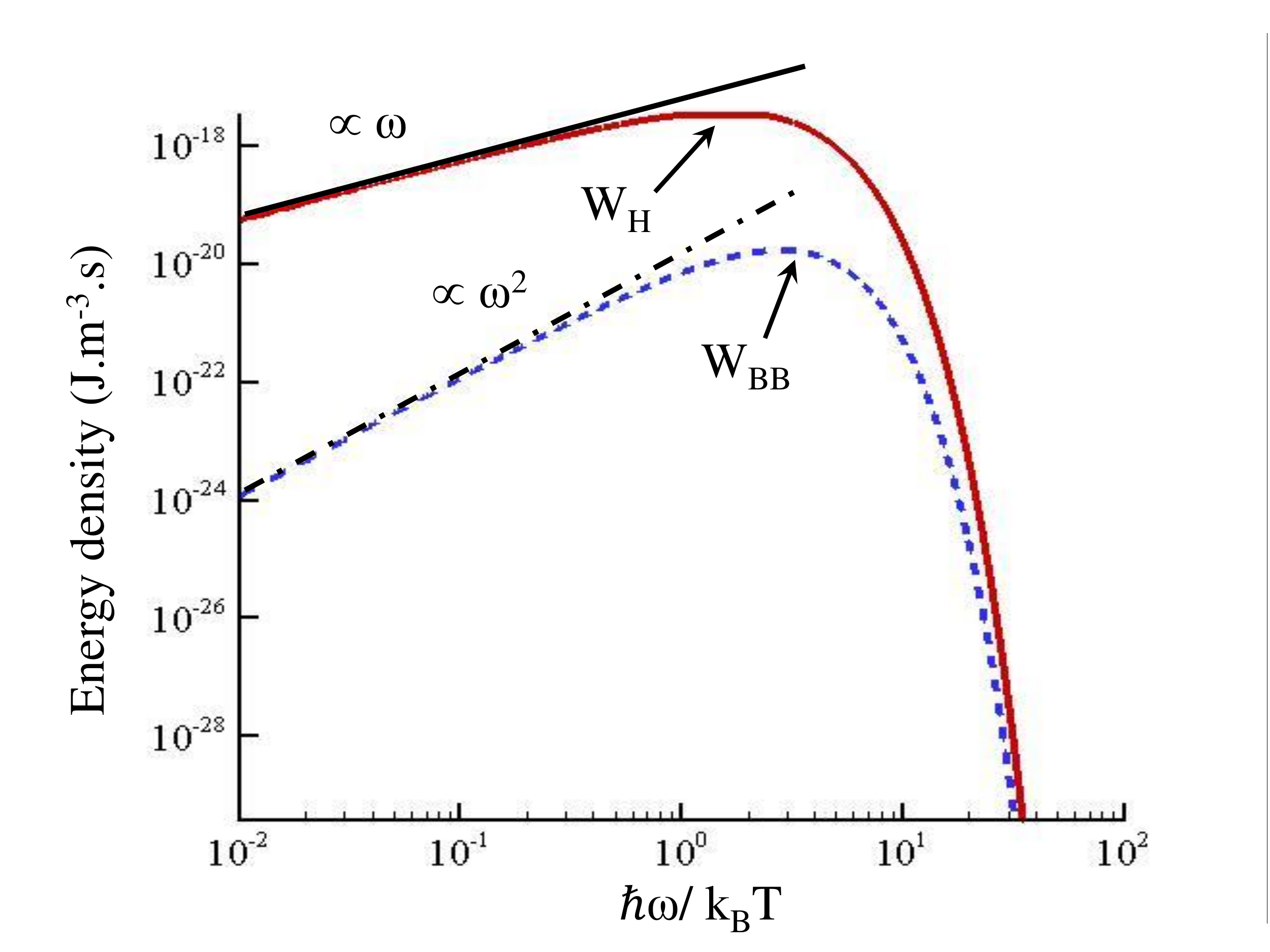, width = 0.5\textwidth}
  \caption{Spectral energy density (in log-log scale) of a type I and type II  hyperbolic BB (red solid line) with $|\epsilon_\parallel \epsilon_\perp|=1$ and $k_{\perp,{\rm max}}=\frac{2\pi}{\Lambda}$ with $\Lambda=100\,{\rm nm}$ at $T=300\,{\rm K}$. This distribution is compared with the classical BB spectrum (blue dashed line). The solid and dashed straight lines show the asymptotic behavior in $\omega$ and $\omega^2$ of the hyperbolic and classical BB spectrum. The arrows indicate Wien's frequencies in both cases.}
  \label{Fig2}
\end{figure}

Another consequence of the linearity of the DOS with respect to $\omega$ inside a hyperbolic 
medium is the spectral shift of Wien's frequency $\omega_{\rm max}$ (resp. wavelength $\lambda_{\rm max}$) 
at which the energy distribution function has its maximum. For both type I and type II hyperbolic 
media we find after a straight forward calculation from relations (9) and (11)  
($k_{\perp, \rm max} \gg \frac{\omega}{c} \sqrt{|\epsilon_\parallel|}$) that this maximum 
is reached when
\begin{equation}
  \frac{\hbar \omega_{\rm max}}{\kb T} = 1.59 \quad \text{or} \quad  \frac{2 \pi l_c}{\lambda_{\rm max}} = 3.92
\end{equation}
whereas for a usual BB $\frac{\hbar \omega_{\rm max}}{\kb T} = 2.82$ and $2 \pi l_c/\lambda_{\rm max} = 4.965$.
Here we have introduced the thermal coherence length $l_c \equiv \frac{\hbar c}{\kb T}$~\cite{MandelWolfBook}. Hence
we see that Wien's frequency is shifted toward smaller values (i.e. the medium appears colder than a classical BB) 
and the maximum vacuum wavelength to larger values (see Fig.~\ref{Fig2}). 

It is now interesting to compare the internal energy of the EMs in a
hyperbolic material with that of a classical BB. From expressions (\ref{Planck_law}) and (\ref{Eq:EnergyDensityHyperbolic})
we immediately get
\begin{equation}
  \frac{U^\re_{\rm I/II}}{U^{\rm p}_{\rm BB}} \approx \sqrt{|\epsilon_\parallel \epsilon_\perp|} (k_{\perp,\rm max} l_c) \frac{\Gamma(3)\zeta(3)}{\Gamma(4)\zeta(4)}.
\label{Eq:EnergyDenshypBB}
\end{equation}
If $\Lambda$ denotes the unit-cell size of our hyperbolic material then $k_{\perp,{\rm max}} = (2 \pi) / \Lambda$, so that
\begin{equation}
  \frac{U^\re_{\rm I/II}}{U^{\rm p}_{\rm BB}} \propto \frac{l_c}{\Lambda}. 
\end{equation}
At a temperature of $300\,{\rm K}$ the coherence length is $l_c = 7.6\,\mu{\rm m}$. The period of realistic
artificial hyperbolic metamaterials is typically larger than $\Lambda \approx 10\,{\rm nm}$. In natural 
hyperbolic materials the unit cell size reduces to the interatomic spacing, i.e.\  $\Lambda \approx 1$ \AA. Hence,
the internal energy of thermal radiation inside a hyperbolic cavity can be 3 to 5 orders of magnitude 
larger than that of a perfect BB. The same is of course also true for the free energy and the entropy. 
This result suggests that the radiative heat flux inside a hyperbolic material is dramatically enhanced compared 
to that of a classical BB.

In order to evaluate the flux radiated by a cavity filled with a hyperbolic medium into a hyperbolic medium and to derive 
Stefan Boltzmann's law we calculate now the Poynting vector in the cavity in the direction of the principal optical axis by assuming, 
for convenience, that the cavity opening (see Fig.~\ref{Fig1}) is along this axis. Using the framework of fluctuational electrodynamics 
theory the ensemble average of the Poynting vector (for any dispersive and dissipative anisotropic medium) reads~\cite{RytovBook1989} 
(Einstein's convention)
\begin{equation}
\begin{split}
  \langle S_\gamma \rangle &= \zeta_{\alpha\beta\gamma} 2 \Re \int_0^\infty \!\!\frac{\rd \omega}{2 \pi} \, \frac{2 \omega^3 \mu_0}{c^2} \mathcal{U}(\omega,T) \\
                           & \quad\times \int_V \!\!\! \rd \mathbf{r}'' \biggl( \mathds{G}^{\rm EE}(\mathbf{r,r''}) \Im (\boldsymbol{\epsilon}) {\mathds{G}^{\rm HE}}^\dagger (\mathbf{r,r''}) \biggr)_{\alpha \beta}.
\end{split}
\end{equation}
Here we have introduced the Levi-Civita tensor $\zeta_{\alpha\beta\gamma}$ and the permeability of vacuum $\mu_0$. 
Note that $\mathds{G}^{\rm HE}(\mathbf{r,r'}) = \frac{1}{\ri \omega \mu_0} \nabla \times \mathds{G}^{\rm EE}(\mathbf{r,r'})$ and that $^\dagger$ denote the conjugate transpose.
In order to determine the heat flux, we assume that the cavity is infinitely large so that we can replace it by a uniaxial halfspace
at a given temperature $T$. Inserting the Green's dyadic~\cite{Weiglhofer1990} and integrating over this halfspace with volume $V$ 
we find after a lengthy calculation (see Ref.~\cite{SupplMat} ) in the lossless limit, the relatively simple expression
\begin{equation}
  \Phi^{\ro/\re} \equiv \langle S_z \rangle = \int_0^\infty \!\! \rd \omega \, \mathcal{U}(\omega,T) \frac{1}{4 \pi^2} \int_0^\infty \!\! \rd k_\perp \, k_\perp \frac{\Re(\gamma_{\ro/\re})^2}{\gamma^2_{\ro/\re}}
\end{equation}
for the mean Poynting vector or heat flux along the surface normal with
\begin{equation}
  \gamma_\ro^2 \equiv \frac{\omega^2}{c^2} \epsilon_\perp - k_\perp^2 \quad \text{and} \quad \gamma_\re^2 \equiv \frac{\omega^2}{c^2} \epsilon_\perp - k_\perp^2 \frac{\epsilon_\perp}{\epsilon_\parallel}.
\end{equation}

Evaluating this expression for the mean Poynting vector for the dielectric uniaxial material, first, we
have
\begin{equation}
    \Phi^{\ro/\re}_{\rm D} = \int_0^\infty \!\!\! \rd \omega\, \mathcal{U}(\omega,T) \frac{\omega^2}{\pi^2 c^3} \frac{c}{4} \frac{1}{2} \begin{Bmatrix} \epsilon_\perp \\ \epsilon_\parallel \end{Bmatrix}.
\end{equation}
For non-dispersive materials this simplifies to
\begin{equation}
    \Phi^{\ro/\re}_{\rm D} = \frac{c}{4} U^{\rm s/p}_{\rm BB} \begin{Bmatrix} \epsilon_\perp \\ \epsilon_\parallel \end{Bmatrix} = \Phi_{\rm BB}^{\rm s/p} \begin{Bmatrix} \epsilon_\perp \\ \epsilon_\parallel \end{Bmatrix}.
\end{equation}
When $\epsilon_\perp = \epsilon_\parallel = 1$ we find again the usual BB result, i.e.\ Stefan-Boltzmann's law. On the other hand, inside a uniaxial material (as inside an
isotropic material) with $\epsilon_\perp > 1$ and  $\epsilon_\parallel > 1$ the radiative heat flux is larger than the BB value, which is a well-known
fact~\cite{Fan}.

In the case of hyperbolic media these results radically change. Before  seeing this, let us first consider the OMs. For $\Phi^{\ro}_{\rm I}$ 
we find of course the same relation as for the dielectric anisotropic material, whereas as a consequence that there do not exist any OMs in a type II 
hyperbolic material we find $\Phi^{\ro}_{\rm II} = 0$. On the contrary, for the EMs we find
\begin{equation}
    \Phi^{\re}_{\rm I} = \int_0^\infty \!\!\!\rd \omega \, \mathcal{U}(\omega,T) \frac{1}{4 \pi^2} \frac{k_{\perp,{\rm max}}^2}{2}
\end{equation}
and 
\begin{equation}
    \Phi^{\re}_{\rm II} = \Phi^{\re}_{\rm I} -  \Phi^{\re}_{\rm D} . 
\end{equation}
Hence,  in the non-dispersive case, where $k_{\perp,{\rm max}} \gg \frac{\omega}{c} \sqrt{|\epsilon_\parallel|}$ we have
\begin{equation}
    \Phi^{\re}_{\rm I} \approx \Phi^{\re}_{\rm II} \approx \frac{k_{\perp,{\rm max}}^2}{8 \pi^2} \frac{\Gamma(2) \zeta(2)}{\hbar}(\kb T)^2.
\end{equation}
In this case, we see that the heat flux is proportional to $T^2$ and not anymore to $T^4$ as in ``classical'' Stefan Boltzmann's law. Comparing this quantity with the classical BB results, we find
\begin{equation}
   \frac{\Phi^{\re}_{\rm I/II}}{\Phi_{\rm BB}^{\rm p}} \approx (k_{\perp,{\rm max}} l_c)^2 \frac{\Gamma(2) \zeta(2)}{\Gamma(4)\zeta(4)}.
\end{equation}
Hence the normalized heat flux is proportional to $(k_{\perp,{\rm max}} l_c)^2$ which is due to the fact that
the heat flux scales like the area of the projection of the isofrequency surface in k-space~\cite{FlorescuEtAl2007,Florescu2009} or like the number
of transversal modes~\cite{BiehsEtAl2010,JoulainPBA2010}, resp.  
We have seen before in Eq.~(\ref{Eq:EnergyDenshypBB}) that the 
ratio of the energy density of a hyperbolic material and that of a BB is only linear in $k_{\perp,{\rm max}} l_c$. This is 
quite astonishing, since for artificial hyperbolic materials with a unit cell size $\Lambda$ of $10\,{\rm nm}$ and for natural 
hyperbolic materials with a unit-cell size  $\Lambda \approx 1\,$\AA \, we can now expect a hyperbolic BB heat flux  
6 to 10 orders of magnitude larger than that of a usual BB at $T = 300 \rm K$. At cryogenic temperatures $l_c$ becomes very 
large so that this ratio can become even much larger. To substantiate this statement let us consider a numerical example: We 
consider an artificial hyperbolic medium made of layered periodic structure of GaN and SiO$_2$ slabs of thickness $5\,{\rm nm}$ each. 
Then, the period is $\Lambda = 10\,{\rm nm}$ so that the theoretical upper limit for the 
wavenumber is $k_{\perp,{\rm max}} \approx 6.28\times 10^8 {\rm m}^{-1}$. If we consider now two blackbodies made 
of and separated by such a material with $T_{\rm mean} = 300\,{\rm K}$ and small temperature difference $\Delta T$ 
then the transferred energy per unit area and temperature (heat transfer coefficient) for the EMs 
is $\Delta \Phi^{\rm e}_{\rm I/II} /\Delta T \approx 8.93 \times10^6\,{\rm W}/{\rm m}^2 {\rm K}$. 
Note that this value for $\Delta \Phi^{\re}_{\rm I/II}/\Delta T$ has to be considered as an upper limit. It is interesting to 
compare this radiative heat flux to the heat conduction by phonons and electrons inside the hyperbolic multilayer 
structure. At ambiant temperature, the thermal conductivity of each 
unit layer, is $\kappa_{\rm GaN} \leq 0.5\, {\rm W/(m K)}$ and $\kappa_{{\rm SiO}_2} \approx 0.4\, {\rm W/(m K)}$~\cite{GaN,SiO2} the effective thermal 
conductivity of the whole structure is about $0.44\, {\rm W/(m K)}$ when assuming that the thermal
resistance of the combined multilayer structure is the averaged sum of the resistances of both materials (i.e.\ Kapitza resistances are neglected). 
Hence the heat transfer coefficient by heat conduction through a hyperbolic multilayer structure of $200\,{\rm nm}$ (20 periods)  
is about $2.2 \times10^6{\rm W}/{\rm m}^2 {\rm K}$. Therefore, a hyperbolic BB can theoretically have a radiative heat flux 
even larger than heat conduction  in weakly conducting composite structures~\cite{Narimanov2011,Narimanov2014}. It is important to note here that when a uniaxial BB radiates 
into vacuum~\cite{Nefedov2011,Biehs,GuoEtAl2012,Biehs2}, the maximum wave number is $k_\perp = \omega/c$. Therefore for dielectric 
uniaxial materials we find, neglecting reflections at the cavity opening, again (if $|\epsilon_\perp| > 1$) the usual BB 
result $\Phi^{\ro/\re}_{\rm D} = \Phi_{\rm BB}^{\rm s/p}$, whereas for hyperbolic 
materials we find in this case 
\begin{equation}
  \Phi^{\ro/\re}_{\rm I} = \Phi_{\rm BB}^{\rm s/p} \quad \text{and} \quad  \Phi^{\ro/\re}_{\rm II} = 0. 
\end{equation}
The type I hyperbolic BB behaves like a perfect BB and the type II hyperbolic BB behaves like a perfect metal or  a``white'' body.

To summarize, we have extended the BB theory to arbitrary uniaxial materials. For dielectric anisotropic media we have seen 
that the thermodynamic properties of the photon gas inside such media are very similar to that of a classical BB. On the other hand, 
when these media are hyperbolic, the spectral energy distibution of radiation is shifted towards frequencies smaller than Wien's frequency 
making these media apparently colder. We have also shown that in contrast to Stefan Boltzmann's law, the heat flux radiated by these media 
depends quadratically on their temperature. Nevertheless, the magnitude of heat flux carried by these media can be several orders of magnitude 
larger than the flux radiated by a classical BB and may even exceed the heat flux carried by conduction in superlattices. Detailed
derivations of the above relations and the underlying assumptions as well as more detailed discussions will be given elsewhere~\cite{SupplMat}.  

%
%

\begin{acknowledgments}
The authors from Hamburg University of Technology gratefully acknowledge financial support from the German Research Foundation (DFG) via SFB 986 "M$^3$", project C1.
\end{acknowledgments}

\end{document}